\documentclass[espcrc2,twoside,fleqn]{article}

\usepackage{epsfig}
\usepackage{espcrc2}

\def\<{\langle}
\def\>{\rangle}
\def\sg{\mbox{\boldmath $\sigma$}}

\title{Composite operators from the operator product expansion: 
what can go wrong?}

\author{Sergio Caracciolo${}^{\rm a}$,
Andrea Montanari\address{Scuola Normale Superiore and INFN, Sezione di Pisa, 
I-56100 Pisa, ITALIA}, and 
Andrea Pelissetto\address{Dipartimento di Fisica and INFN, Sezione di Roma I,
Universit\`a degli Studi di Roma ``La Sapienza'', I-00185 Roma, ITALIA}}

\begin{document}

\begin{abstract}
The operator product expansion is used to compute the matrix elements of 
composite renormalized operators on the lattice. We study the product of two 
fundamental fields in the two-dimensional $\sigma$-model and discuss 
the possible sources of systematic errors. The key problem turns out
to be the violation of asymptotic scaling.
\end{abstract}

\maketitle

The Operator Product Expansion
\begin{eqnarray}
A(x;\mu)B(-x;\mu) \sim \sum_C W_{AB}^C(x;\mu)C(0;\mu)
\label{ShortDistance}
\end{eqnarray} 
is widely thought to hold beyond perturbation theory.

The use of Eq. (\ref{ShortDistance}) in 
lattice simulations \cite{DeltaI,NostroLat98,Schierolz} is still in its 
infancy.
In Ref. \cite{DeltaI} it was suggested to use Eq. (\ref{ShortDistance}) in 
order to compute renormalized matrix elements.
The OPE approach
consists in the following steps: 
one computes the (matrix element of the) l.h.s. of 
Eq. (\ref{ShortDistance}), then renormalizes $A$ and $B$ 
in some scheme, and finally obtains
(the matrix element of) $C$ through a fit, 
using some perturbative approximation of the Wilson coefficients.

The main sources of systematic errors in this approach are the following:
\renewcommand{\theenumi}{(\alph{enumi})}
\begin{enumerate}
\item finite-size effects and corrections to scaling, i.e. lattice artifacts; 
\label{ScalingProblem}
\item ``power-correction effects'' which are due to the fact that we 
truncate the expansion (\ref{ShortDistance}) to some finite order in $x^2$;
\label{HTProblem}
\item corrections to asymptotic scaling which must be taken in account since
the Wilson coefficients in Eq. (\ref{ShortDistance}) have to be substituted by
the first few terms of their perturbative expansion.
\label{AsymptoticProblem}
\end{enumerate}
Errors of type \ref{ScalingProblem} are widely studied and 
do not need more explanations. Here we shall focus on errors of type 
\ref{HTProblem} and \ref{AsymptoticProblem}.
 
The use of Eq. (\ref{ShortDistance}) in a continuum scheme, for which only a 
perturbative computation of the Wilson coefficients is available, poses
a restriction on the operators which can be obtained in this approach.
Only the operators of lowest dimension for each spin sector, 
i.e. those of lowest twist, can be computed using 
Eq. (\ref{ShortDistance}). Higher-twist operators 
give rise to systematic errors 
of order $O(x^2)$. Moreover, because of statistical
errors, only the operators of low dimension can be reliably extracted:
the remaining ones are strongly subleading in the region of validity of 
the OPE.

Problem \ref{AsymptoticProblem} can be stated in a cleaner way if we 
get rid of the scale dependence which is introduced in this approach 
somehow artificially through the Wilson coefficients. One can rewrite 
Eq. (\ref{ShortDistance}) by making use of renormalization-group invariant 
operators defined as follows:
\begin{eqnarray}
\lefteqn{ Q^{RGI}(x)\equiv Q(x;\mu)/F_Q(g(\mu))\; ,} \\
\lefteqn{ F_Q(g)  \equiv  g^{\frac{\gamma^Q_0}{\beta_0}}
\exp\left\{\int_0^{g} \left[\frac{\gamma^Q(x)}{\beta(x)}-
\frac{\gamma^Q_0}{\beta_0 x}\right]dx\right\} \; . }
\end{eqnarray}
The Wilson coefficients obviously become $\mu$-independent and their 
general perturbative form is
\begin{eqnarray}
W^{RGI}(\Lambda x) = g(\Lambda x)^{\frac{\gamma_0^W}{\beta_0}}
\sum_{k=0}^{\infty} c_k  g(\Lambda x)^k \; ,
\label{RGIPT}
\end{eqnarray}
where $\Lambda$ is the ``lambda parameter'' of the theory.
The use of a truncation of Eq. (\ref{RGIPT}) introduces systematic errors 
of order $O(\log^{-k}(\Lambda x))$.
Notice, however, that this approach allows to compute directly 
``infinite-energy'' quantities (i.e. the renormalization-group invariant 
matrix elements) which 
are of interest in phenomenological applications. Errors 
of order $\log^{-k} (\Lambda x)$ arise also in the widely 
used ``non-perturbative 
renormalization method'' \cite{NonPerturbativeMethod} 
in which perturbation theory
is used to ``evolve'' the renormalization constants computed at some 
energy scale achievable on the lattice up to high energies.

We have considered several products of operators for the $O(N)$ nonlinear
$\sigma$-model in two dimensions, with lattice action
\begin{eqnarray}
S(\sg) \equiv \frac{1}{2g_L}\sum_{x,\mu}(\partial_{\mu}\sg)_x^2 \; ,
\end{eqnarray}
where $\sg\in S^{N-1}$, $(\partial_{\mu}f)_x \equiv f_{x+\mu}-f_x$, and $N=3$.
Here we shall refer to the following (respectively scalar and symmetric) 
products of fundamental fields:
\begin{eqnarray}
\lefteqn{\sg(x)\cdot\sg(-x)\sim W_0(x) +O(x^2)\; ,}
\label{ProdottoScalare}\\
\lefteqn{\sigma^a(x)\sigma^b(-x)+(a\leftrightarrow b)-
\mbox{trace}\sim}\nonumber\\ 
&&\sim W_2(x)\left[\sigma^a\sigma^b-\mbox{trace}\right](0)+O(x^2) \; .
\label{ProdottoSimmetrico}
\end{eqnarray}
The Monte Carlo data presented refer to two lattices: the first one of 
size $L\times T=128\times 256$ and correlation length 
$(am)^{-1} = 13.632(6)$ ($m$ is the mass gap); the second with 
$L\times T=256\times 512$ and correlation length $(am)^{-1} = 27.094(43)$.

\begin{figure}[t]
\epsfig{file=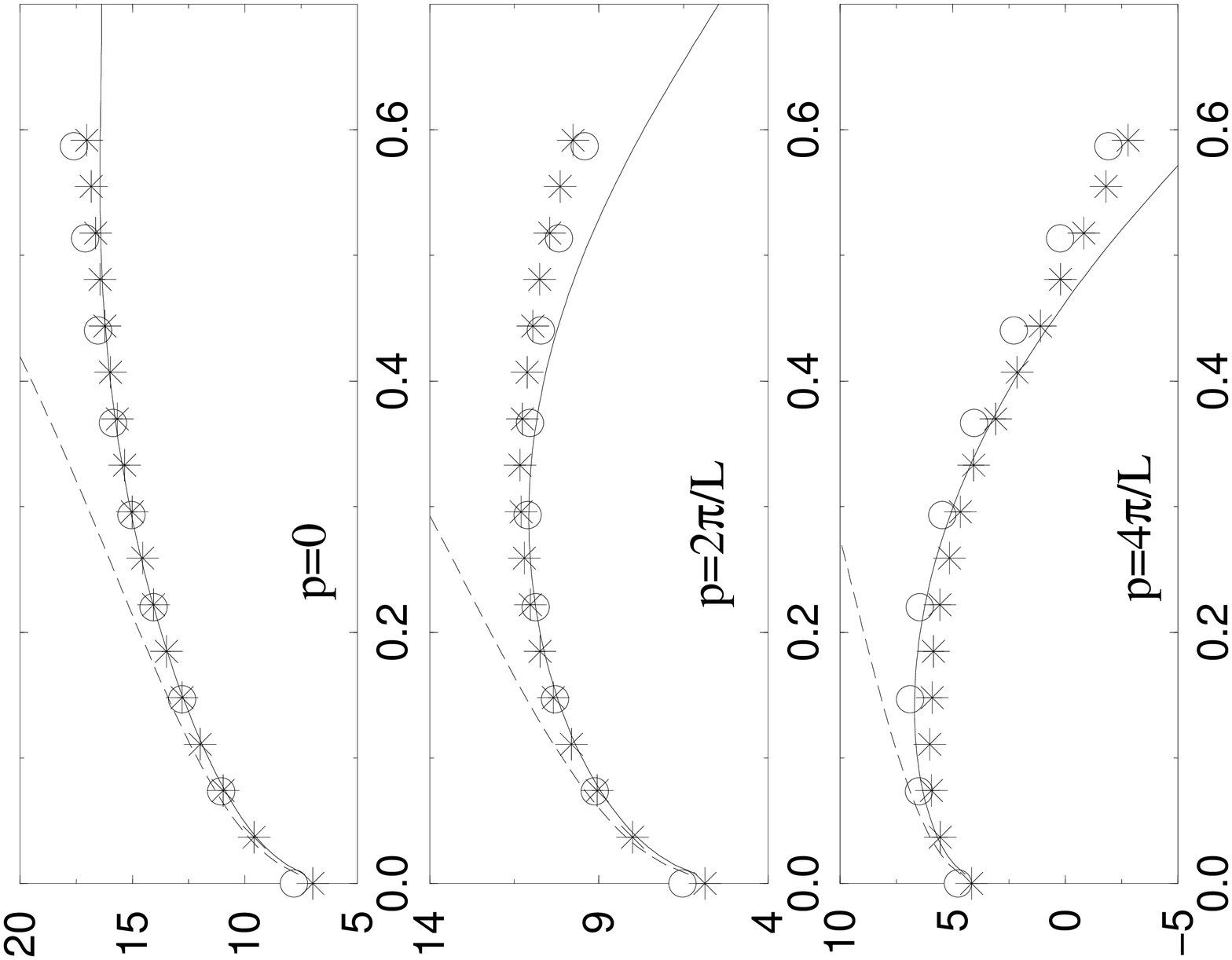,angle=-90,width=1.0\linewidth}
\caption{The product (\ref{ProdottoSimmetrico}) on the slice $x_0=0$.
Circles are the data at $(ma)^{-1}\simeq 13.632$ and stars at 
$(ma)^{-1}\simeq 27.094$ (properly rescaled). 
Continuous and dashed lines are the fitting 
curves respectively with and without $O(x^2)$ terms.}
\label{FunzioneDiCorrelazione}
\vspace{-0.6cm}
\end{figure}

The expectation value of the product (\ref{ProdottoSimmetrico}) 
between states of momentum $p$ is shown in 
Fig. \ref{FunzioneDiCorrelazione} for the two different lattices:
corrections to scaling, i.e. errors of type \ref{ScalingProblem},
are completely under control in our simulations. \\
Our general procedure consists in choosing a truncation of the expansion
(\ref{ShortDistance}) and in using it to fit the data in the region 
$\rho< |x| < R$. The results are independent of $\rho$ for 
$1.5\ a\le\rho\le 3\ a$.
We use the stability of the fit with respect to the truncation and to 
the choice of $R$ as a criterion to distinguish whether 
errors of type \ref{HTProblem} and \ref{AsymptoticProblem} are relevant or not.

The quality of the fits obtained is well represented by the curves shown in
Fig. \ref{FunzioneDiCorrelazione}, where the two-loop expression was used
for the first term in the expansion (\ref{ProdottoSimmetrico}) and 
the tree-level form for the terms of order $O(x^2)$.

An additional conclusion can be drawn from Fig. \ref{FunzioneDiCorrelazione}:
in order to describe the symmetric product (\ref{ProdottoSimmetrico}) up to
distances $2x\sim m^{-1}, p^{-1}$, it is necessary (and almost sufficient) 
to include terms of order $O(x^2)$. However this does not mean that 
the terms of order $x^2$ with the same symmetry of the leading one 
--- the higher-twist terms --- 
can be obtained from the fit. Indeed, 
their matrix elements extracted from the fits 
are very unstable with respect to changes of $R$.
\begin{figure}
\epsfig{file=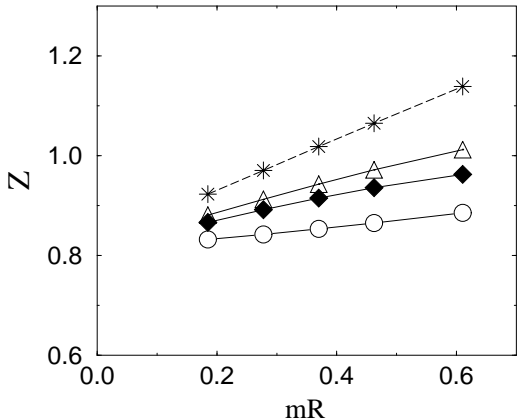,angle=-90,width=1.0\linewidth}
\caption{The field-renormalization constant as computed 
with different truncations of the OPE.
{}From top to bottom: one loop, two loops, three loops, three loops plus 
$O(x^2)$ corrections.}
\label{Z}
\vspace{-0.6cm}
\end{figure} 

In order to test the stability of the fit, we considered  
the vacuum expectation value of the product (\ref{ProdottoScalare}), that is
the two-point function. Here the renormalization constant of the 
field is a fit parameter. The results are reported in  Fig. \ref{Z} 
and refer to the lattice with correlation length
$(ma)^{-1}\simeq 27.094$; we used $\overline{\mu}a \simeq 10$.
As a manifestation of asymptotic freedom, the various curves shrink when 
$R\to 0$. Their $R$-dependence becomes weaker as more terms of the 
perturbative expansion are included. Nevertheless, even if 
we use the three-loop Wilson 
coefficient, we do not obtain a value of $Z$ independent of $R$ 
within the statistical errors, as it should be in the asymptotic-scaling 
regime. Two remarks are in order here: first, it is notoriously 
difficult to reach asymptotic scaling in the $O(3)$ nonlinear 
$\sigma$-model \cite{Asymptotic}; second, we are studing a small effect which 
could be negligible in QCD applications with respect to other errors.

Note that the addition of $O(x^2)$ terms, which, in this case, have 
spin $0$ and are therefore higher twists, seems to improve the situation, 
making the curve flatter. However, this must be regarded as a spurious 
effect. Higher twists are simply mimicking the contribution of higher orders 
in perturbation theory: otherwise, the curves obtained without including 
them should show a $(mR)^2$ behavior.
{}From Fig. \ref{Z} we can estimate the systematic error 
due to the use of two- or three-loop Wilson coefficients to fit the data in this
range of $mR$: the error is approximately $5\%$. \\
\begin{figure}
\epsfig{file=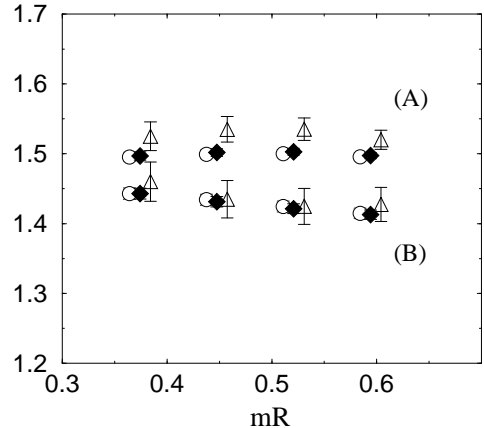,angle=-90,width=1.0\linewidth}
\caption{The expectation value $\<p|\sigma^a\sigma^b-\mbox{trace}|p\>$ 
with momentum: $p=0$ (circles); $p=2\pi/L$ (filled diamonds); 
$p=4\pi/L$ (triangles). 
Two different fits are reported: $O(x^2)$ corrections are not considered 
in fit (A), while they are included in fit (B).}
\label{ElementoDiMatrice}
\vspace{-0.6cm}
\end{figure}
We report in Fig. \ref{ElementoDiMatrice} the results of the fit 
presented in Fig. \ref{FunzioneDiCorrelazione}. Each type of truncation,
including or not $O(x^2)$ corrections, gives a reasonable, that is $p$ 
independent and (almost) $R$ independent, answer. The difference between them
should be interpreted as a violation of asymptotic scaling and, indeed,
 it is of 
the same magnitude of the systematic error estimated for the renormalization 
constant of the field.
  
We thank R.~Petronzio for useful discussions.

\end{document}